\documentclass[preprint,12pt]{elsarticle}
\usepackage{hyperref}
\usepackage{bm}
\usepackage[normalem]{ulem}
\usepackage{color}
\usepackage{geometry}
\usepackage[caption=false,listofformat=subsimple,
  labelformat=simple]{subfig}
 \usepackage{graphicx}

\usepackage{amssymb}
\usepackage{amsmath}

\journal{Physica A}

\begin{document}

\begin{frontmatter}

\title{Decorated Clusters and Geometrical Frustration in Cluster Spin
  Glass: A Random Graph Approach}

\author[ufrgs]{S. G. Magalhães}
\ead{sgmagal@gmail.com}

\author[ufms]{F. M. Zimmer}
\ead{fabiozimmer@gmail.com}

\author[ufrgs]{R. Erichsen Jr.}
\ead{rubem@if.ufrgs.br}

\affiliation[ufrgs]{organization={Instituto de Física,
    Universidade Federal do Rio Grande do Sul},
  city={Porto Alegre},
  postcode={91501-970},
  state={RS},
  country={Brazil}
}

\affiliation[ufms]{organization={Instituto de Física,
    Universidade Federal do Mato Grosso do Sul},
  city={Campo Grande},
  state={MS},
  country={Brazil}}

\begin{abstract}

We develop a theory to investigate how geometrically frustrated clusters that become decorated affect the Cluster Spin Glass phase. The cluster structure is assumed to be a tetrahedron composed of Ising spins with z-anisotropy placed at its vertices that interact antiferromagnetically. We consider the probability $1-p_J$ of finding an impurity at a vertex of the tetrahedron that interacts ferromagnetically with the remaining elements inside the tetrahedron. An intercluster disorder is added as a random Gaussian interaction. The order parameters are obtained using the sparse random graph technique, which introduces the connectivity of the network of clusters as a controllable parameter in the theory. We examine changes that occur in the Cluster Spin Glass phase as a function of $p_J$ and $c$, in addition to the antiferromagnetic intracluster couplings $J_1$. For intermediate values of $p_J$, unexpected results appear. Even when some clusters contain a ferromagnetic impurity, there will still be robust geometric frustration effects in the cluster network. However, the $p_J$ threshold for this to occur depends on connectivity. Conversely, below this threshold, reduced GF effects favor the reappearance of the CSG phase. Furthermore, the Curie-Weiss temperature $\Theta_W$ has a gradual change of signal, indicating that the effects of the impurities extend to the paramagnetic phase.

\end{abstract}


\end{frontmatter}

\section{Introduction}

The frustration effects on spin systems have such complexity that they took the corresponding theory down a fertile path of conceptual innovation \citep{Lacroixbook}. For example, the theory proposed to deal with frustration and quenched disorder has generated such an overpowering conceptual framework \citep{FischerSG,mezard1987spin} that it spilled into other areas of knowledge as distinct as, for example, biology \citep{Pagnani_2009} and information theory and combinatorial optimization \citep{Montanari}. Interestingly, the combination of frustration and disorder can also boost new platforms for magnetic materials containing innovative functionalities. One of the most promising possibilities comes from clusters of spins. A possible area of interest for this platform would be, for instance, molecular nanomagnetism \citep{Schnack2010, Furrer2013}. Furthermore, recently, the inclusion of Geometric Frustration (GF) in the description of cluster system having Cluster Spin Glass (CSG) phase, besides its intriguing scientific issues, has acquired technological interest. That is due to its potential to enhance the magnetocaloric effect \citep {Zimmer2023, Kumar2024, Mishra2024}.  However, little consideration has been given to describe how the decorated cluster, defined as those with an inner disorder, interfere with GF, affecting, as a result, the CSG phase.

The high degeneracy introduced by geometrical frustration can make the system extremely sensitive to small perturbations. A common source of perturbation is disorder, which is inevitable due to imperfections in real physical systems. In geometrically frustrated systems, even a very small amount of disorder can have a significant impact, often driving the system into a glassy state. For instance, results reported for the pyrochlore lattice claim that a spin-glass state can be found in the presence of very low levels of  quenched disorder  \cite{PhysRevLett.98.157201, PhysRevLett.114.247207}. Furthermore, the presence of small frustrated clusters can enhance the effects of disorder, acting as an important mechanism for the freezing process and the stabilization of glassy order in weakly disordered frustrated systems \cite{Zimmer2022, Schmidt2015416}. 

In spin systems with clusters, the non-trivial ergodicity breaking of the SG can be described within the so-called SG Cluster model \citep{soukoulis0,soukoulis1}. This theory has a random infinite-range intercluster couplings and an intracluster part where the spins are bonded with ferromagnetic (FE) or antiferromagnetic (AF) couplings. The random intercluster part can be handled at the mean-field level using the replica method \citep{PhysRevLett.35.1792,Parisi_19802,Parisi_19803}. As a result, the combination of infinite-ranged random couplings with locally ones, allow the stabilization of a Cluster Spin Glass (CSG) phase. Moreover, the intracluster part of the SG Cluster model requires a solver for each cluster structure. Thus, GF can be introduced by choosing tetrahedral, triangular, or kagome clusters with AF couplings.  For frustrated triangular and kagomé clusters, the CSG phase is robust even for infinitesimal random interaction between clusters \citep{Zimmer2014A, SchmidtScripta2015,Mateus2016, silveira2021}. These previous works deal with locally ordered clusters. In contrast, here we deal with clusters that are decorated with a ferromagnetic impurity, introducing local disorder. Would the inner disorder of the cluster suppress the effects of GF? Would be the CSG phase destabilized or reinforced?

To answer properly the above questions, we propose going beyond the mean-field approximation. Indeed, this approximation corresponds to the rather artificial limit of fully connected clusters. Instead, in this work, we adopt an alternative approach using sparse random graphs (SRG) \citep{mezard-parisi87, Monasson, Montanari2015}, where the connectivity appears as a random variable \citep{Metz2019,rubem2021}. As a result, it becomes a controllable parameter within the theory.

Therefore, we aim to investigate how clusters of spins, that were initially geometrically frustrated, affect the stability of the CSG phase when they become decorated. We start assuming a lattice of random interacting clusters of Ising spins as in the SG Cluster model. However, instead of using the mean-field approximation, we use SRG. Thus, the cluster system is transformed into a network of clusters with connectivity following a Poissonian distribution. Moreover, we initially consider clusters with a tetrahedral structure composed of Ising spins with z-anisotropy placed at its vertices that interact antiferromagnetically.  Then, a disorder is introduced that allows a given vertex of the tetrahedron to be randomly occupied by an impurity with probability $1-p_J$ with the spin impurity interacting ferromagnetically with the remaining spins of the tetrahedron.

We use the replica method \citep{Monasson} to deal with the disordered intercluster random interaction.  In other words, the SRG replicated partition function is obtained within the replica symmetry solution (RS) \citep{silveira2021}. However, since the RS solution is unstable, to mark this instability, the CSG boundary phase line is also obtained by the two-replica method \citep{tworeplica}.

The paper is organized as follows: the model and theoretical framework are presented in Section \ref{model}. Section \ref{results} presents the numerical results and discussions. Conclusions and further remarks are given in Section \ref{conclusion}.

\section{The Model and Theoretical Framework}
\label{model}

We consider a random network composed of $N_c$ interacting clusters,
each of them composed of $p$ spins assuming unidirectional Ising spin
states with z-axis anisotropy
$\sigma_{i\mu}=\pm 1/2$, where $i=1\dots p$ is the local index and
$\mu=1\dots N_c$ is the cluster index. From now on, this random
network of interacting clusters will be called the Cluster Network
(CN). The clusters interact through their total spin,
\begin{align}
    \sigma_\mu=\sum_{i=1}^p\sigma_{i\mu}\,.
    \label{totalspin}
\end{align} 
The Hamiltonian is given by
\begin{align} 
H(\boldsymbol{\sigma}{,}\{J_{\mu\nu}\}{,}\{J_{ij}\})=
-\frac{1}{\sqrt{c}} \sum_{\mu<\nu,\nu}c_{\mu\nu}J_{\mu\nu}
\sigma_{\mu}\sigma_{\nu} +
\sum_{\mu}H_{0}(\boldsymbol{\sigma}_{\mu}{,} \{J_{ij}\})\,,
\label{hamiltonian} 
\end{align}
where $\boldsymbol{\sigma}=\{\sigma_{i\mu}\}$ represents the state of
the whole system, $\boldsymbol{\sigma}_\mu$ represents the state of
cluster $\mu$ and $J_{\mu\nu}$ are intercluster couplings, randomly
chosen from a gaussian distribution with zero average and unitary
variance, which is equivalent to fix the energy scale in units of
$J_{\mu\nu}$ variance. The elements of the matrix connectivity
$c_{\mu\nu}$ are chosen according the binary distribution
\begin{align}
  P(c_{\mu\nu})=\frac{c}{N_c}\delta_{c_{\mu\nu}{,}1} +
  \Big(1-\frac{c}{N_c}\Big)\delta_{c_{\mu\nu}{,}0}\,.
  \label{connect}
\end{align}

The local Hamiltonian $H_0$ is given by
\begin{align}
  H_0(\boldsymbol{\sigma}_\mu{,}\{J_{ij}\})=-\sum_{i<j,j}
  J_{ij}\sigma_{i\mu}\sigma_{j\mu}\,,
  \label{intra}
\end{align}
where the $J_{ij}$ are intracluster couplings, to be specified later.

We apply the replica method, assuming that the disorder averaged free
energy density (i.e., the free energy per cluster) can be written as
\begin{align}
    f(\beta)=-\lim_{\substack{N_{cl}\rightarrow\infty \\ n\rightarrow
        0}} \frac{1}{\beta nN_c} \ln\langle {Z^n}\rangle\,,
    \label{free}
\end{align}
where
\begin{equation}
  Z^n=\sum_{\boldsymbol{\sigma}^1\cdots\boldsymbol{\sigma}^n}
  \mathrm{e}^{-\beta \sum_{\alpha=1}^n
    H(\boldsymbol{\sigma}^\alpha{,}\{J_{\mu\nu}\}{,}\{J_{ij}\})}\,
    \label{part1}
\end{equation}
is the replicated partition function, being $\alpha$ the replica
index.

Introducing the Hamiltonian, Eq. (\ref{hamiltonian}), in the
replicated partition function and averaging over the $c_{\mu\nu}$ we
obtain, as the leading order in $c/N$,
\begin{align}
  \label{fpart}
    \langle Z^{n} \rangle =
    \sum_{\boldsymbol{\sigma}^1\cdots\boldsymbol{\sigma}^n}
    \Big\langle\exp\Big[& -\beta\sum_{\alpha\mu}
      H_{0}(\boldsymbol{\sigma}_\mu^\alpha{,}\{J_{ij}\}) + \beta
      h_e\sum_{\alpha\mu}\sigma_\mu^\alpha \\ & + \frac{c}{2N_c}
      \sum_{\mu\neq \nu}\Big\langle
      \mathrm{e}^{\frac{\beta}{\sqrt{c}}J_{\mu\nu}\sum_{\alpha}
        \sigma_\mu^\alpha\sigma_\nu^\alpha} -1
      \Big\rangle_{J_{\mu\nu}}\Big]\Big\rangle_{\{J_{ij}\}}\,.
    \nonumber
\end{align}
To reduce the problem of a single cluster, the cluster spin variables
are withdrawn from the inner exponential by introducing the auxiliary
variables $s_\alpha=\sum_{i=1}^ps_i^\alpha$. By doing so, the
partition function becomes
\begin{align}
\label{fpartn}
    \langle Z^{n} \rangle =
    \sum_{\boldsymbol{\sigma}^1\cdots\boldsymbol{\sigma}^n}
    \Big\langle\exp\Big[& -\beta\sum_{\alpha\mu}
      H_{0}(\boldsymbol{\sigma}_\mu^\alpha{,}\{J_{ij}\}) + \beta
      h_e\sum_{\alpha\mu}\sigma_\mu^\alpha \\ & + \frac{c}{2N_c}
      \sum_{\mu\neq \nu}\sum_{\mathbf{s}
        \mathbf{s}'}\delta_{\mathbf{s}
        \boldsymbol{\sigma}_{\mu}}\delta_{\mathbf{s}'\boldsymbol{\sigma}_{\nu}}
      \Big\langle\mathrm{e}^{\frac{\beta J}{\sqrt{c}}
        \sum_{\alpha}s_\alpha
        s'_\alpha}-1\Big\rangle_J\Big]\Big\rangle_{J_{\{ij\}}}\,.
    \nonumber
\end{align}
Technical details are left to the \Ref{appendix}. The main outcome of
the finite connectivity replica calculation is the saddle-point
equation
\begin{align}
W(\mathbf{h}) = \sum_k P_k \int\prod_{l=1}^k
d\mathbf{h}_l\,W(\mathbf{h}_l) \Big\langle\prod_{i=1}^4\delta\Big(h_i
- \sum_l\phi_i(\mathbf{h}_l{,}J_l{,}\{J_{ij}^l\})\Big)
\Big\rangle_{\{J_l\},\{J_{ij}^l\}}\,, \label{RS9}
\end{align}
where $W(\mathbf{h})$ is the distribution of the $p$-component local
fields $\mathbf{h}$, the $P_k$ are poissonian weights and the
$\phi_i(\mathbf{h}_l{,}J_l{,}\{J_{ij}^l\})$ are functions defined in
the \Ref{appendix}. The method to solve this equation will be
explained below.

The RS free-energy density is obtained by introducing the RS {\sl
  Ansatz} in Eq. (\ref{freen3}), resulting
\begin{align}
  \nonumber
    f(\beta) = \frac{c}{2\beta}\int d\mathbf{h}\,d\mathbf{h}' &
    W(\mathbf{h})W(\mathbf{h}')
    \Bigg\langle\ln\dfrac{\bar{\chi}(\mathbf{h}{,}\mathbf{h}'{,}J{,}
      \{J_{ij}\}{,}\{J_{ij}'\})}{\chi(0{,}\mathbf{h}{,}0{,}\{J_{ij}\})
      \chi(0{,}\mathbf{h}{,}0{,}\{J_{ij}'\})}\Bigg\rangle_{J,\{J_{ij}\},\{J_{ij}'\}}
    \\ & \nonumber - \frac{1}{\beta}\sum_k P_k\int\prod_l
    d\mathbf{h}_l W(\mathbf{h}_l) \Bigg\langle \ln\sum_{\mathbf{s}}
    e^{\beta(- H_0(\mathbf{s}{,}\{J_{ij}\}) + h_e s)}\\ &
    \qquad\qquad\times\prod_l
    \frac{\chi(s{,}\mathbf{h}{,}J_l{,}\{J_{ij}\})
    }{\chi(0{,}\mathbf{h}{,}0{,}\{J_{ij}\})}\Bigg\rangle_{\{J_l\},\{J_{ij}\}}
    \,,
\end{align}
where
\begin{align}
\nonumber \bar{\chi}(\mathbf{h}{,}\mathbf{h}'{,}J{,}\{J_{ij}\} &
          {,}\{J_{ij}'\}) =\sum_{\mathbf{s}\mathbf{s}'}
          \exp\Big[-\beta H_0(\mathbf{s}{,}\{J_{ij}\}) \\ & +
            \beta\mathbf{h}\cdot\mathbf{M}(s) - \beta
            H_0(\mathbf{s}'{,}\{J_{ij}'\}) +
            \beta\mathbf{h}'\cdot\mathbf{M}(s') +\beta
            \frac{J}{\sqrt{c}}ss'\Big]\,.
\end{align}

Other relevant observables are the CSG order parameter and the
temperature-dependent square of the total spin of each cluster in the
CN given, respectively, by
\begin{align}
    q=\int d\mathbf{h}\, W(\mathbf{h})\langle s(\mathbf{h})\rangle^2
  \label{qea}
\end{align}
and
\begin{align}
    Q=\int d\mathbf{h}\, W(\mathbf{h})\langle
    s^2(\mathbf{h})\rangle\,,
  \label{sg}
\end{align}
where
\begin{align}
    \langle s(\mathbf{h})\rangle=\Bigg\langle\dfrac{\sum_{\mathbf{s}}
      s\exp\big[-\beta H_0(\mathbf{s}{,}\{J_{ij}\}) +
        \beta\mathbf{h}\cdot\mathbf{M}(s)\big]} {\sum_{\mathbf{s}}
      \exp\big[-\beta H_0(\mathbf{s}{,}\{J_{ij}\}) +
        \beta\mathbf{h}\cdot\mathbf{M}(s)\big]}\Bigg\rangle_{\{J_{ij}\}}
    \,,
  \label{sigmaav}
\end{align}
and
\begin{align}
    \langle
    s^2(\mathbf{h})\rangle=\Bigg\langle\dfrac{\sum_{\mathbf{s}}
      s^2\exp\big[-\beta H_0(\mathbf{s}{,}\{J_{ij}\}) +
        \beta\mathbf{h}\cdot\mathbf{M}(s)\big]} {\sum_{\mathbf{s}}
      \exp\big[-\beta H_0(\mathbf{s}{,}\{J_{ij}\}) +
        \beta\mathbf{h}\cdot\mathbf{M}(s)\big]}\Bigg\rangle_{\{J_{ij}\}}
    \,.
  \label{sigma2av}
\end{align}

The freezing temperature $T_f$ coincides with the de Almeida-Thouless
(AT) line \citep{AT}, which signals the stability of the RS
solution. In the finite connectivity approach, the AT line can be
found using the two-replica method \citep{tworeplica}. To do so, we
calculate the two-replica joint distribution
$W(\mathbf{h}_l{,}\mathbf{h}'_l)$ through the saddle-point equation
\begin{align}
\label{AT1}
W&(\mathbf{h}{,}\mathbf{h}') = \sum_k P_k \int\prod_{l=1}^k
d\mathbf{h}_l\,W(\mathbf{h}_l{,}\mathbf{h}'_l)\\ &
\times\Big\langle\prod_{i=1}^4\Big[\delta\Big(h_i -
  \sum_l\phi_i(\mathbf{h}_l{,}J_l{,}\{J_{ij}^l\})\Big)\delta\Big(h_i'
  - \sum_l\phi_i(\mathbf{h}'_l{,}J_l{,}\{J_{ij}^l\})\Big)\Big]
\Big\rangle_{\{J_l\},\{J_{ij}^l\}}\,. \nonumber
\end{align}
The overlap between the two replicas is given by
\begin{align}
    q'=\int d\mathbf{h}d\mathbf{h}'W(\mathbf{h},\mathbf{h}')\langle
    s(\mathbf{h})\rangle \langle s(\mathbf{h}')\rangle\,.
    \label{overlaptworep}
\end{align}
The RS solution is stable if $q=q'$ and unstable otherwise.

\section{Results and Discussion}
\label{results}

\begin{figure}
\centering \includegraphics[width=3cm,clip]{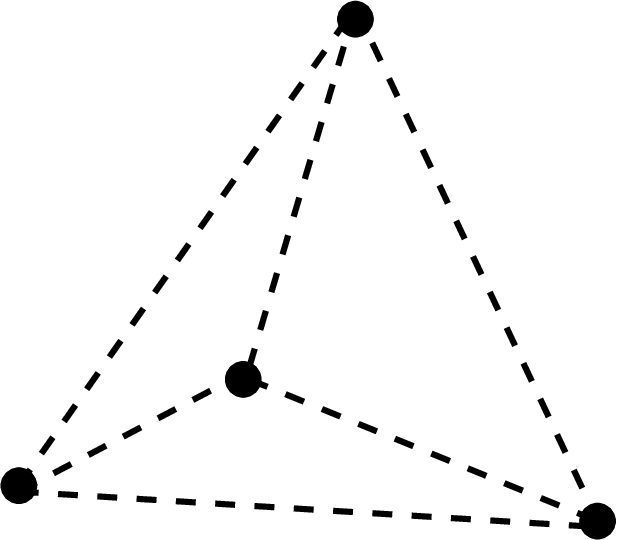}\hspace{3cm}
\includegraphics[width=3cm,clip]{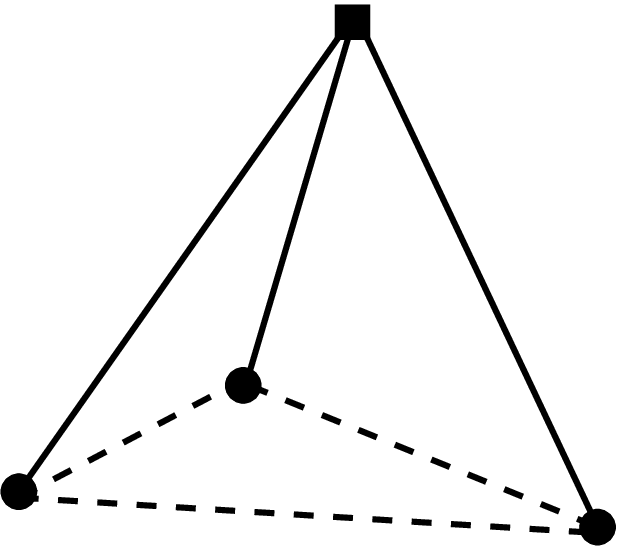}

(a)\hspace{5cm} (b)
    \caption{Tetrahedron with solid lines representing ferromagnetic
      (FE) bonds and dashed lines representing antiferromagnetic (AF)
      ones.  Circles represent the element A and squares represent the
      impurity B.  (a) Ordered, $p_J=1$ homogeneous tetrahedron with
      $J_1=J_{AA}<0$.  (b) Disordered, $p_J=0$ tetrahedron with
      $J_1=J_{AA}<0$ and $J_1=J_{AB}>0$.  }
    \label{fig:clusters}
\end{figure}

We start with a cluster with tetrahedral geometry. Then, one has four
vertices occupied with elements A that have antiferromagnetic (AF)
bonds among them. The disorder inside the cluster is introduced by
assuming the probability $1-p_J$ of having one site occupied by an
impurity B that has ferromagnetic (FE) bonds with elements A. Thus,
the bond disorder inside the cluster is expressed as
\begin{align}
    P_1(\mathrm{E})=p_J \ \delta_{\mathrm{E,A}} + (1-p_J) \ \delta_{\mathrm{E,B}}\,,
\end{align}
where $P_1(\mathrm{E})$ is the probability that one vertex of the
tetrahedron be occupied by element E $\in$ \{A{,B}\}. The remaining
vertices are occupied by element A.  To simplify the discussion and
avoid increasing the number of parameters, we choose $J_{ij}=J_{AA}$
if the vertices $i$ and $j$ are occupied by the same element $A$ and
$J_{ij}=J_{AB}$ if sites $i$ and $j$ are occupied by distinct
elements, $A$ and $B$.  The configurations of bonds in the tetrahedron
for $p_J=1$ and $0$ are shown in Fig. \ref{fig:clusters} with
$J_{AA}=J_1=-J_{AB}$.  In both scenarios, the tetrahedron is
geometrically frustrated. The difference is the total spin of the
tetrahedron $S_T$ minimizing the energy and the number of frustrated
bonds. Thus, $S_T=0$ and $1$ for $p_J=1$ and $0$, respectively.

\begin{figure}
\centering \includegraphics[width=12cm]{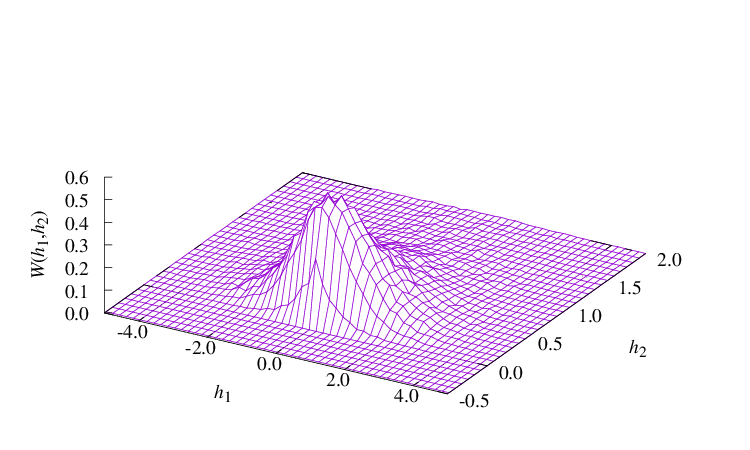}\\ (a)

\includegraphics[width=12cm]{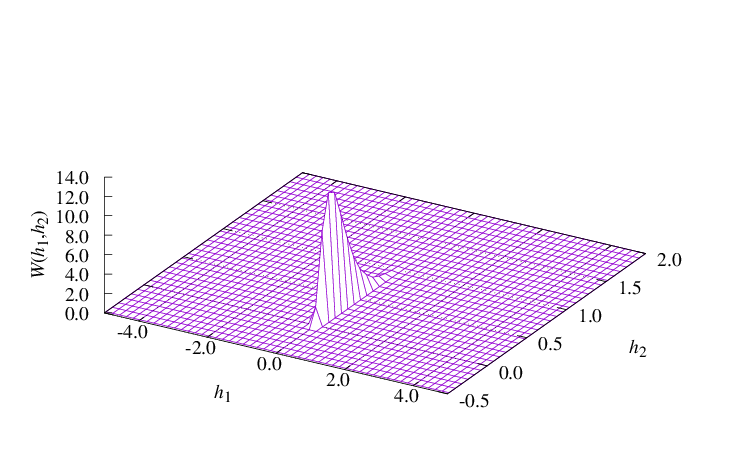}\\ (b)
\caption{(a) Marginal distributions $W(h_1{,}h_2)$ for $c=8$, $p_J=1$,
  $J_1=-1$ and $T=0.5$. (b) The same, but for $T=1.5$. }
\label{fig:distr}
\end{figure}

We employ a recursive algorithm to solve the fixed-point
Eq. (\ref{RS9}). A population $\mathcal{N}$ of 4-dimensional vector
fields is randomly created. Then for each step, in sequence: (i) an
integer $k$ is chosen randomly according to a Poisson distribution;
(ii) $k$ 4-components fields are chosen randomly from the population;
(iii) the summation appearing in the Dirac's delta function in
Eq. (\ref{RS9}) is calculated, for each field component and the
results are stored in a further field, also randomly chosen from the
distribution. This procedure runs till it converges to a stable
distribution. The size $\mathcal{N}$ reflects the accuracy of the
procedure. The larger the $\mathcal{N}$, the larger the convergence
time $t$. After some experience we are convinced that the choice
$\mathcal{N}=100{,}000$ and $t=20{,}000{,}000$, assumed throughout
this work, allows to acceptable results. We assume the temperature and
$J_1$ are given in units of $J$. As an illustration,
Fig. \ref{fig:distr} reveals the resulting marginal field
distributions $W(h_1{,}h_2)$ for $c=8$, $p_J=1$ and $J_1=-1$. In
Fig. \ref{fig:distr}(a) the temperature is $0.5$. A symmetric $h_1$
distribution represents a CSG phase with a finite order parameter. In
Fig. \ref{fig:distr}(b) the temperature is $1.5$. This time, the
marginal $W(h_1)$ is a delta-shaped distribution corresponding to null
order parameter, typical of a paramagnetic (PM) phase.

\begin{figure}
    \centering
    \includegraphics[width=5cm]{Qq_2_1.eps}\hspace{1cm}
    \includegraphics[width=5cm]{Qq_4_1.eps}
    
    \hspace{0.5cm}(a)\hspace{5.5cm} (b)

    \vspace{1cm}
    \includegraphics[width=5cm]{Qq_8_1.eps}\hspace{1cm}
    \includegraphics[width=5cm]{Qq_12_1.eps}

    \hspace{0.5cm}(c)\hspace{5.5cm} (d)

   \caption{ Order parameters $Q$ and $q$ versus temperature $T$ for
     $p_J=1$; $J_1=-2$ (solid), $J_1=-4$ (dotted) and $J_1=-6$
     (dashed).  (a) $c=2$; (b) $c=4$; (c) $c=8$; (d) $c=12$.}
    \label{fig:QqpJ1}
\end{figure}

\begin{figure}
    \centering
    \includegraphics[width=5cm]{Qq_2_.5.eps}\hspace{1cm}
    \includegraphics[width=5cm]{Qq_4_.5.eps}
    
    \hspace{0.5cm}(a)\hspace{5.5cm} (b)

    \vspace{1cm}
    \includegraphics[width=5cm]{Qq_8_.5.eps}\hspace{1cm}
    \includegraphics[width=5cm]{Qq_12_.5.eps}

    \hspace{0.5cm}(c)\hspace{5.5cm} (d)

   \caption{ Order parameters $Q$ and $q$ versus temperature $T$ for
     $p_J=0.5$; $J_1=-2$ (solid), $J_1=-4$ (dotted) and $J_1=-6$
     (dashed).  (a) $c=2$; (b) $c=4$; (c) $c=8$; (d) $c=12$.}
    \label{fig:QqpJ.5}
\end{figure}

\begin{figure}
    \centering
    \includegraphics[width=5cm]{Qq_2_0.eps}\hspace{1cm}
    \includegraphics[width=5cm]{Qq_4_0.eps}
    
    \hspace{0.5cm}(a)\hspace{5.5cm} (b)

    \vspace{1cm}
    \includegraphics[width=5cm]{Qq_8_0.eps}\hspace{1cm}
    \includegraphics[width=5cm]{Qq_12_0.eps}

    \hspace{0.5cm}(c)\hspace{5.5cm} (d)

   \caption{ Order parameters $Q$ and $q$ versus temperature $T$ for
     $p_J=0$; $J_1=-2$ (solid), $J_1=-4$ (dotted) and $J_1=-6$
     (dashed).  (a) $c=2$; (b) $c=4$; (c) $c=8$; (d) $c=12$.}
    \label{fig:QqpJ0}
\end{figure}

Figs. \ref{fig:QqpJ1}-\ref{fig:QqpJ0} show the behavior of the order
parameters $q$ and $Q$ versus the temperature.  We start the
discussion with the undecorated tetrahedron, $p_J=1$, where the
conditions to the appearing of a CSG phase are met only for low
connectivity values, e.g. $c=2$ and $c=4$, and weak AF intracluster
coupling, e.g. $-J_1\leq 2$. The order parameter $Q$ shows a
non-monotonic behavior, while the order parameter $q$ decreases
monotonously to zero at a $c$-dependent $T_f$. In the absence of a
low-temperature CSG phase for $c=8$ and $c=12$, $Q$ is zero at $T=0$
and then increases monotonically as the temperature increases.  The
presence of a CSG phase in this condition can be understood by means
of the fluctuations induced by the disordered intercluster
couplings. These fluctuations, being larger for smaller $c$ values,
can stabilize a CSG phase for $J_1=-2$ and $c=2$ or $c=4$, but not for
larger $c$ or $|J_1|$. The way the fluctuations induces the CSG phase
deserves an explanation. Consider $c=2$. This means that several
clusters will have only one neighbor and they will strongly influence
each other. In this case, the intercluster coupling may favor
tetrahedron states $S_T\neq 0$. Conversely, if $c$ is large, the
probability to favor states $S_T\neq 0$ will be smaller, since the
cluster neighborhood will be more equally distributed. Furthermore,
the long-range couplings are normalized to $1/\sqrt{c}$ (see
Eq. \ref{hamiltonian}).

If $0<p_J<1$, part of the tetrahedrons in the CN are decorated. In
contrast, for $p_J=0$, the CN is composed with every tetrahedron
having a ferromagnetic impurity, called from now on, fully
decorated. Now, for both cases, the CSG phase does not result from
fluctuations, but from the interactions between the decorated
tetrahedrons, which increase with $c$. This time, contrary to $p_J=1$,
is the large connectivity and the weakening of GF effects in the CN
that favor the CSG phase. The case $p_J=0.5$ is shown in
Fig. \ref{fig:QqpJ.5}.  Except for $c=2$ and $J_1=-6$, the freezing
temperature $T_f$ is always finite.  However, the precise location of
$T_f$ depends on $c$ and $J_1$ values. A similar picture is observed
for $p_J=0$ (see Fig. \ref{fig:QqpJ0}). This time, the presence of the
ferromagnetic impurity in every tetrahedron favors larger $Q$ and $q$
states overall and higher $T_f$ values.

\begin{figure}
    \centering
    \includegraphics[width=7cm]{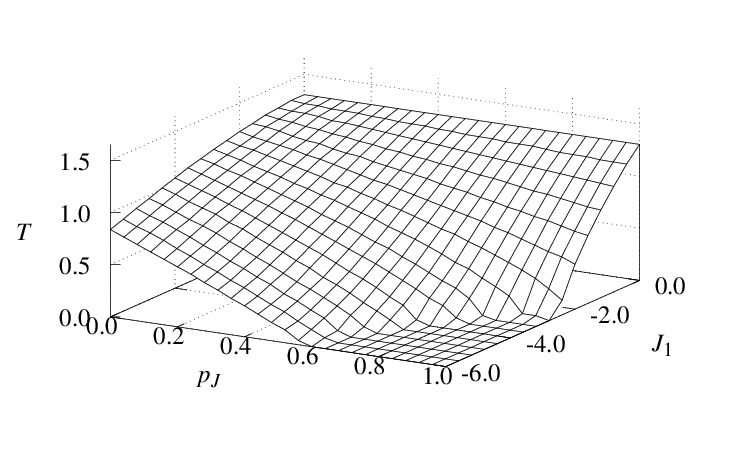}
    \includegraphics[width=7cm]{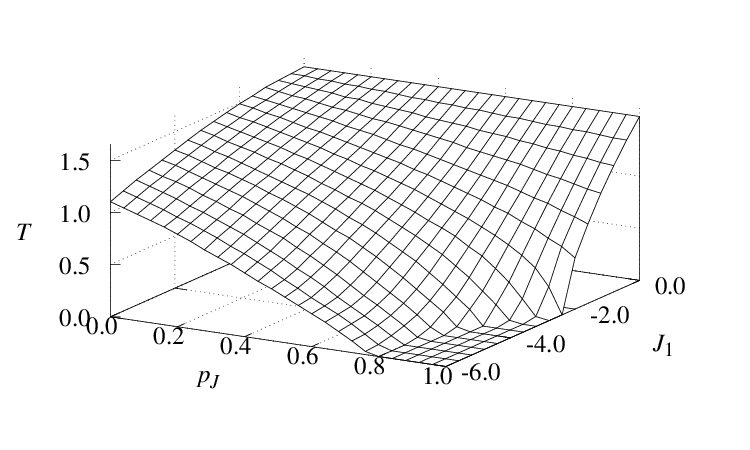}

    \vspace{-0.5cm} (a) \hspace{7cm} (b)

    \includegraphics[width=7cm]{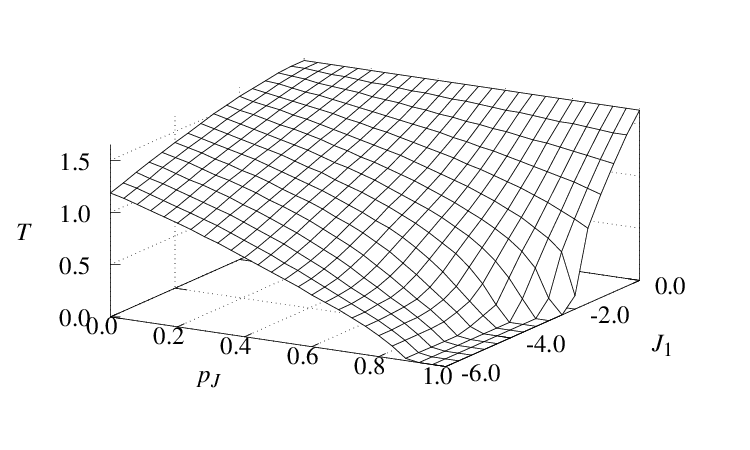}
    \includegraphics[width=7cm]{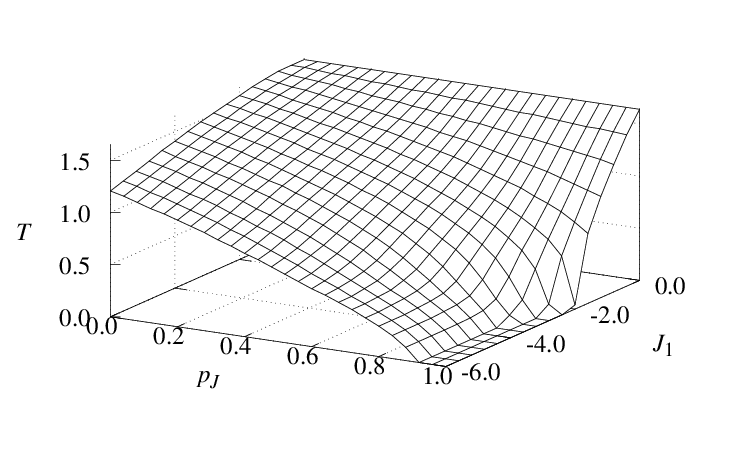}

    \vspace{-0.5cm} (c) \hspace{7cm} (d)

    \caption{Freezing temperature $T_f$ as a function of $p_J$ and
      $J_1$ for some representative connectivity values.  (a) $c=2$;
      (b) $c=4$; $c=8$; (d) $c=12$. }
    \label{fig:TpJJ1}
\end{figure}

To summarize our findings on the order parameters behavior,
Fig. \ref{fig:TpJJ1} displays phase diagrams $T$ vs $p_J$ vs $J_1$
elaborated from solutions for $q$ and $Q$ in
Eqs. (\ref{qea})-(\ref{sg}). The CN connectivity is ranging from $c=2$
to $c=12$. For $p_J=0$, the freezing temperature $T_f$ is finite for
any value of $J_1$ and $c$. However, it is worthwhile to remark that
$T_f$ decreases for smaller values of $c$ and larger $-J_1$. In
contrast, for instance, for $J_1=-3$, there is suppression of the CSG
phase at $p_J^{*}\approx 0.9$ for any value of $c$.  Moreover, with
larger fixed $-J_1$, the role of $c$ is amplified, i.e., a smaller $c$
further decreases the value of $p_J^*$ where the CSG is
suppressed. Thus, reinforcing the hypothesis that GF effects in the CN
are also strengthened by $c$ besides the intracluster coupling $J_1$,
leading to a decreasing $p_J^*$. This possibility is discussed below
examining the entropy.
\begin{figure}
    \centering{
    \includegraphics[width=5cm]{s_2_1.eps}\hspace{1cm}
    \includegraphics[width=5cm]{s_4_1.eps}
    
    \hspace{0.5cm}(a)\hspace{5.5cm} (b)

    \vspace{1cm}
    \includegraphics[width=5cm]{s_8_1.eps}\hspace{1cm}
    \includegraphics[width=5cm]{s_12_1.eps}

    \hspace{0.5cm}(c)\hspace{5.5cm} (d) }
   \caption{ Entropy density $s$ versus temperature $T$ for $p_J=1$;
     $J_1=-2$ (solid), $J_1=-4$ (dotted) and $J_1=-6$ (dashed).  (a)
     $c=2$; (b) $c=4$; (c) $c=6$; (d) $c=12$. Thick (thin) lines
     indicates RS stable (unstable) solutions.}
    \label{fig:spJ1}
\end{figure}

\begin{figure}
    \centering{
    \includegraphics[width=5cm]{s_2_.5.eps}\hspace{1cm}
    \includegraphics[width=5cm]{s_4_.5.eps}
    
    \hspace{0.5cm}(a)\hspace{5.5cm} (b)

    \vspace{1cm}
    \includegraphics[width=5cm]{s_8_.5.eps}\hspace{1cm}
    \includegraphics[width=5cm]{s_12_.5.eps}

    \hspace{0.5cm}(c)\hspace{5.5cm} (d) }
   \caption{ Entropy density $s$ versus temperature $T$ for $p_J=0.5$;
     $J_1=-2$ (solid), $J_1=-4$ (dotted) and $J_1=-6$ (dashed).  (a)
     $c=2$; (b) $c=4$; (c) $c=6$; (d) $c=12$. Thick (thin) lines
     indicate RS stable (unstable) solutions.}
    \label{fig:spJ.5}
\end{figure}

\begin{figure}
    \centering{
    \includegraphics[width=5cm]{s_2_0.eps}\hspace{1cm}
    \includegraphics[width=5cm]{s_4_0.eps}
    
    \hspace{0.5cm}(a)\hspace{5.5cm} (b)

    \vspace{1cm}
    \includegraphics[width=5cm]{s_8_0.eps}\hspace{1cm}
    \includegraphics[width=5cm]{s_12_0.eps}

    \hspace{0.5cm}(c)\hspace{5.5cm} (d) }
   \caption{ Entropy density $s$ versus temperature $T$ for $p_J=0$;
     $J_1=-2$ (solid), $J_1=-4$ (dotted) and $J_1=-6$ (dashed).  (a)
     $c=2$; (b) $c=4$; (c) $c=6$; (d) $c=12$. Thick (thin) lines
     indicates RS stable (unstable) solutions.}
    \label{fig:spJ0}
\end{figure}

Next, we discuss the entropy density $s$ as a function of the
temperature. Fig. \ref{fig:spJ1} shows $s$ versus $T$ for with $p_J=1$
and several values of CN connectivities and coupling constant
$J_1$. The independence of the entropy relative to $c$, except for
$J_1=-2$, demonstrates that the CN itself is governed by the
geometrically frustrated undecorated tetrahedrons. So, the CN entropy
density converges to a residual entropy density per cluster given as
$s_0=\ln 6$ as $T\rightarrow 0$. Only for $J_1=-2$, where there is a
CSG phase at low temperature, the entropy becomes
$c$-dependent. However, if $c$ is large enough the entropy density
also converges to its residual value $s_0$, demonstrating that the FG
effects become stronger with increasing $c$ associated with the
suppression of the CSG phase.

The cases of $p_J=0.5$ and $p_J=0$ are shown in Figs. \ref{fig:spJ.5}
and \ref{fig:spJ0}, respectively. For $p_J=0.5$, only for small $c$
and sufficiently large $J_1$ (see Fig \ref{fig:spJ.5}-a) the
geometrically frustrated tetrahedrons do prevail in the CN. That leads
to the convergence of the entropy density to $s_0$ and the suppression
of the CSG phase, as shown in Fig \ref{fig:QqpJ.5}-a.  In other words,
in contrast to the undecorated case, for $p_J=0.5$, the GF effects
become weaker with increasing $c$. For the fully decorated case, there
is no trace of FG effects. On the contrary, the CSG phase is
stabilized in any case shown in Fig. \ref{fig:spJ0}. However, the
behavior of the entropy density at $T<T_f$ is unreliable due to the
instability of the RS solution.  The RSB range, marked using the two
replica method, is indicated by thin lines in the entropy density
curves.

\begin{figure}
    \centering
    \includegraphics[width=7cm]{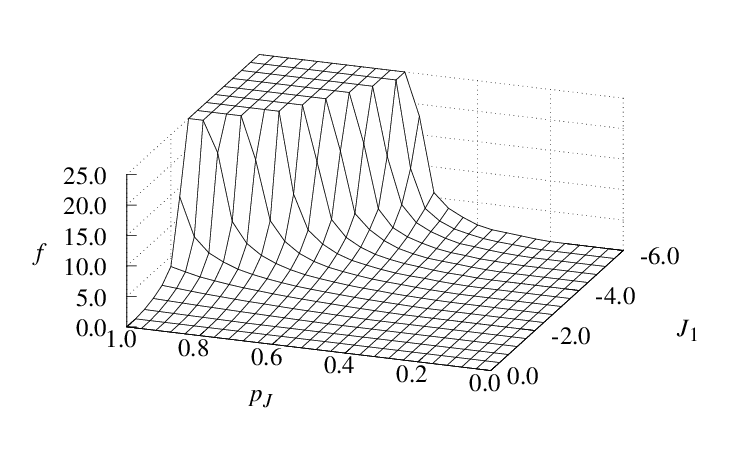}
    \includegraphics[width=7cm]{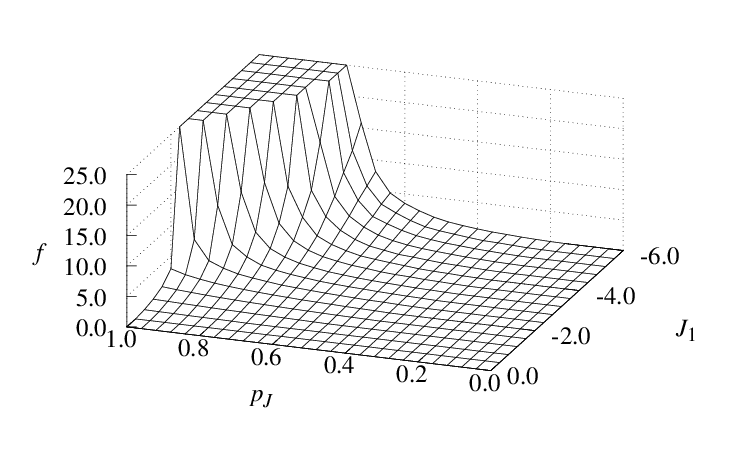}
    
    \hspace{-0.5cm}(a)\hspace{7cm} (b)
    
    \includegraphics[width=7cm]{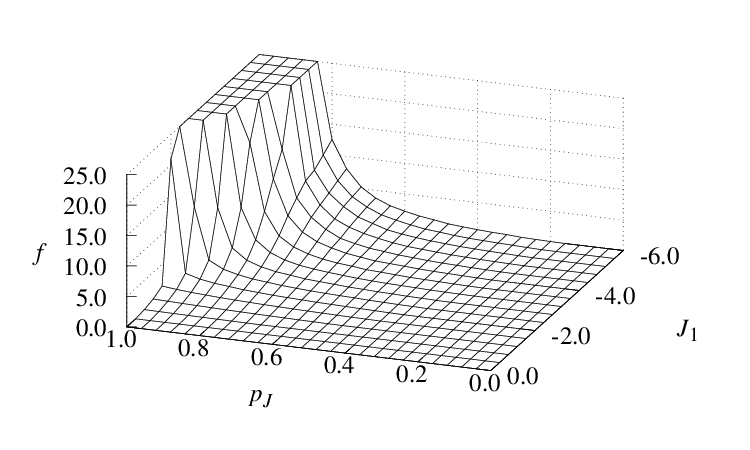}
    \includegraphics[width=7cm]{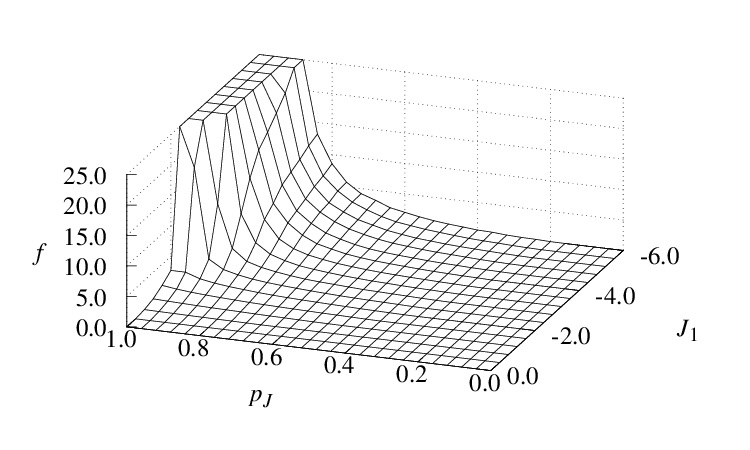}

    \hspace{-0.5cm}(c)\hspace{7cm} (d)

    \caption{Measure of frustration $f$ vs $p_J$ and $J_1$. The
      connectivity is (a) $c=2$; (b) $c=4$; (c) $c=8$; (d) $c=12$.}
    \label{fig:TpJ-4}
\end{figure}

In Fig. \ref{fig:TpJ-4}, a measure of frustration, given by
$f_{p}=-\Theta_W/T_f$ \citep{ramirez,ramirez2} where $\Theta_W$ is the
Curie-Weiss temperature is shown as a function of $p_J$ and $J_1$.
The Curie-Weiss temperature is obtained using the usual procedure of
extrapolating the high-temperature behavior of the inverse of
susceptibility $\chi^{-1}$ to $T=0$.  In Fig. \ref{fig:TpJ-4}, there
are parts of the plane $p_J$ versus $J_1$ where $f_p$ diverges or is
null. In the first case (cutoff at $f=25$), $T_f$ goes to zero, i.e.,
the CSG phase is suppressed (see Fig. \ref{fig:TpJJ1}) by GF
effects. In the second, it is assumed that $f_p=0$ for $\Theta_W>0$.
Fig. \ref{fig:TpJ-4} shows that for $J_1$ sufficiently large, the
divergence of $f_p$ occurs even when the CN is partially decorated.
However, except for the undecorated case, the region where $f_p$
diverges shrinks as $c$ increases. Interestingly, there is also a
gradual change in the sign of $\Theta_W$ as the CN changes from fully
to partially decorated independently of the connectivity.

\section{Concluding remarks}
\label{conclusion}

In this work, we developed a theory to describe a network of
geometrically frustrated clusters when a portion or even all of them
have a defect. We call these decorated clusters. In our theory, defect
means specifically the following two steps. Firstly, we assume a
tetrahedral cluster structure where all vertices are occupied by
elements interacting antiferromagnetically. Then, a ferromagnetic
impurity is included with a probability $1-p_J$ of occupying some
vertex in the tetrahedron.  Moreover, we use Sparse Random Graph (SGR)
technique which allows a description beyond the mean-field
approximation. In other words, the connectivity among the clusters is
a controllable parameter. In this case, we are dealing with a cluster
network.  There is also a random interaction between clusters that
helps to stabilize a Cluster Spin Glass (CSG) phase.  Thus, we draw
our conclusions from changes that occur in the CSG phase.

In the undecorate case $p_J=1$, e.g. none cluster has a ferromagnetic
impurity, the GF suppresses the CSG phase except for low connectivity
and small antiferromagnetic couplings. In this limit, a mechanism
related to the connectivity fluctuations intervenes to stabilize the
CSG phase. In the opposite limit $p_J=0$, the CSG is always
stabilized. There is no trace of GF impact on the CSG phase. However,
there are some unexpected consequences for $0<p_J<1$.  For example,
even when part of the clusters has a ferromagnetic impurity, there
will still be robust GF effects in the cluster network. This is the
case as long as the antiferromagnetic couplings within the cluster are
sufficiently strong. Nonetheless, the threshold of $p_J$ for this to
happen depends on $c$. 
Conversely, below this threshold, reduced GF effects favor the reappearance of the CSG phase. Indeed, the presence of ferromagnetic impurities provides a way to replace geometrically frustrated clusters with those in which frustration arises from competition between FE and AF intracluster interactions. These competing interactions produce the same degree of intracluster degeneracy as the fully geometrically frustrated case. However, the degenerate local (cluster) spin states differ between the two situations: geometrically frustrated clusters produce zero magnetic moments, while FE-AF competition can result in nonzero cluster magnetic moments. The intracluster disorder given by $p_J$ would interpolate between the two situations. We also find a gradual change in the sign of the Curie-Weiss temperature $\Theta_W$ as more clusters have ferromagnetic impurities, indicating that the accumulation of defects in the cluster network also manifests itself in the paramagnetic phase.

Lastly, we highlight that the theory presented in this work can be
applied to clusters with other geometries. We are currently
investigating this problem with cluster structures such as kagomé and
triangular \citep{Sampathkumaran2024} as well as using other types of
disorder within the cluster. Another direction of investigation in which our type of approach may be useful is artificial spin glass systems consisting of Ising-spin nanomagnets arranged in a Hopfield-type network that have recently been observed \citep{Saccone2022}.

\appendix
\section{Derivation of the saddle-point equation}
\label{appendix}

We continue with Eq. (\ref{fpartn}) and introduce the order function
\begin{align}
    P(\mathbf{s})
    =\frac{1}{N_c}\sum_\mu\delta_{\mathbf{s}\boldsymbol{\sigma}_\mu} \label{orderfunc}
\end{align}
as the probability to find a replica state vector $\mathbf{s}$ and its
conjugated order function $\hat{P}(\mathbf{s})$ to obtain
\begin{align}
    \nonumber \big\langle Z^n\big\rangle =
    \sum_{\boldsymbol{\sigma}^1\cdots\boldsymbol{\sigma}^n} &
    \Big\langle\int\prod_{\mathbf{s}}dP(\mathbf{s})
    d\hat{P}(\mathbf{s})\exp\Big\{- \beta\sum_{\alpha\mu}
    H_0(\boldsymbol{\sigma}_\mu^\alpha{,}\{J_{ij}\}) \\ + \beta h_e &
    \sum_{\alpha\mu}\sigma_\mu^\alpha + \sum_{\mathbf{s}}
    \hat{P}(\mathbf{s})P(\mathbf{s}) - \frac{1}{N_c}\sum_{\mathbf{s}}
    \hat{P}(\mathbf{s})\sum_\mu\delta_{\mathbf{s}\boldsymbol{\sigma}_\mu} \label{fpart2}
    \\ & + \frac{cN_c}{2} \sum_{\mathbf{s}\mathbf{s}'}
    P(\mathbf{s})P(\mathbf{s}') \Big\langle\mathrm{e}^{\frac{\beta
        J}{\sqrt{c}}\sum_\alpha s_\alpha s'_\alpha}-1\Big\rangle_J
    \Big\}\Big\rangle_{\{J_{ij}\}}\,.  \nonumber
\end{align}
Summing over the cluster variables and changing $\hat{P}(\mathbf{s})$
to $N_c\hat{P}(\mathbf{s})$, the partition function becomes
\begin{align}
  \label{fpart3} \big\langle Z^n \big\rangle = \int \prod_{\mathbf{s}}
  dP(\mathbf{s}) d\hat{P}(\mathbf{s}) & \exp N_c
  \Big\{\sum_{\mathbf{s}} \hat{P}(\mathbf{s})P(\mathbf{s}) \\ +
  \ln\sum_{\mathbf{s}} \Big\langle \exp\Big[ -\beta & \sum_\alpha
    H_0(\mathbf{s}^\alpha{,}\{J_{ij}\}) + \beta h_e\sum_\alpha
    s_\alpha -\hat{P}(\mathbf{s})\Big] \Big\rangle_{\{J_{ij}\}}
  \nonumber \\ & \quad\quad + \frac{c}{2} \sum_{\mathbf{s}\mathbf{s}'}
  P(\mathbf{s})P(\mathbf{s}') \Big\langle\mathrm{e}^{\frac{\beta J}
    {\sqrt{c}} \sum_\alpha s_\alpha s'_\alpha} - 1\Big\rangle_J \Big\}
  \nonumber \,.
\end{align}
In the limit $N_c\rightarrow\infty$, this integral can be evaluated by
the saddle-point method. Eliminating the $\hat{P}(\mathbf{s})$ through
the saddle-point equations, we can write the free energy
(Eq. \ref{free}) as
\begin{align}
  \label{freen3} f(\beta) =-\lim_{n\rightarrow 0} \frac{1}{\beta n}\mathrm{Extr}
  \Big\{-\frac{c}{2} \sum_{\mathbf{s}\mathbf{s}'} & P(\mathbf{s})
  P(\mathbf{s}') \Big\langle\mathrm{e}^{\frac{\beta
      J}{\sqrt{c}}\sum_\alpha s_\alpha s'_\alpha} - 1\Big\rangle_J
  \\ \nonumber + \ln\sum_{\mathbf{s}} & \Big\langle \exp\Big[ -
    \beta\sum_\alpha H_0(\mathbf{s}^\alpha{,}\{J_{ij}\}) + \beta
    h_e\sum_\alpha s_\alpha \\ & + c\sum_{\mathbf{s}'} P(\mathbf{s}')
    \Big\langle\mathrm{e}^{\frac{\beta J}{\sqrt{c}} \sum_\alpha
      s_\alpha s'_\alpha} - 1
    \Big\rangle_J\Big]\Big\rangle_{\{J_{ij}\}} \Big\}\,, \nonumber
\end{align}
where $\mathrm{Extr}\{\cdot\}$ means the extremum relative to
$P(\mathbf{s})$. Applying this condition we obtain a self consistent
equation for $P(\mathbf{s})$:
\begin{align}
  \nonumber
  P(\mathbf{s})=\frac{1}{\mathcal{N}}\Big\langle\exp\beta\sum_\alpha
  \Big[& - H_0(\mathbf{s}^\alpha{,}\{J_{ij}\}) + h_e s_\alpha\Big]
  \Big\rangle_{\{J_{ij}\}} \\ & \times\exp\Big[c\sum_{\mathbf{s}'}
    P(\mathbf{s}') \Big\langle\mathrm{e}^{\frac{\beta
        J}{\sqrt{c}}\sum_\alpha s_\alpha s'_\alpha} -
    1\Big\rangle_J\Big]\,, \label{saddle3}
\end{align}
with $\mathcal{N}\equiv\sum_{\mathbf{s}}P(\mathbf{s})$ is a
normalization factor.  We focus on the replica symmetric (RS) solution
of Eq. (\ref{saddle3}), by which the order function remains unchanged
under replica permutation. This is expressed by the RS {\sl Ansatz}
\begin{align}
  P(\mathbf{s})=\int d\mathbf{h}\, W(\mathbf{h}) \Bigg\langle
  \dfrac{\exp\beta\sum_\alpha
    \big[-H_0(\mathbf{s}^\alpha{,}\{J_{ij}\}) + h_e s_\alpha +
      \mathbf{h} \cdot \mathbf{M}(s_\alpha)\big]}
        {\big\{\sum_{\mathbf{s}}\exp\beta\big[-
            H_0(\mathbf{s}{,}\{J_{ij}\}) + h_e s +
            \mathbf{h}\cdot\mathbf{M}(s)\big]\big\}^n}
        \Bigg\rangle_{\{J_{ij}\}}\,, \label{RSansatz}
\end{align}
where $\mathbf{h}$ and $\mathbf{M}(s)\equiv(s{,}s^2{,}\dots{,}s^p)$
are $p$ components vectors and $W(\mathbf{h})$ is a $p$ component
distribution of local fields. Introducing the {\sl Ansatz} in
Eq. (\ref{saddle3}) and expanding the exponential, summing over the
spin variables and rearranging terms we have, unless for the
normalization factor $\mathcal{N}$,
\begin{align}
  \label{RS5} P(\mathbf{s}) = \Big\langle &\exp\beta\sum_\alpha
  \big[- H_0(\mathbf{s}^\alpha{,}\{J_{ij}\}) + h_e s_\alpha\big]
  \Big\rangle_{\{J_{ij}\}} \\ & \times\sum_k P_k \prod_{l=1}^k \int
  d\mathbf{h}_l\,W(\mathbf{h}_l) \Bigg\langle\frac{\exp\sum_\alpha
    \sum_s\delta_{ss_\alpha}
    \ln\chi(s{,}\mathbf{h}_l{,}J_l{,}\{J_{ij}^l\})}
  {\chi^n(0{,}\mathbf{h}_l{,}0{,}\{J_{ij}^l\})}
  \Bigg\rangle_{J_l,\{J_{ij}^l\}}\,, \nonumber
\end{align}
where $p_k=e^{-c}c^k/k!$ is a poissonian weight and
\begin{align}
  \chi(s{,}\mathbf{h}{,}J{,}\{J_{ij}\})=\sum_{\mathbf{s}'}\exp\beta
  \Big[- H_0(\mathbf{s}'{,}\{J_{ij}\}) + h_e s' +
    \mathbf{h}\cdot\mathbf{M}(s') + \frac{J}{\sqrt{c}}ss'\Big]\,.
\end{align}

The total spin of the tetrahedral, 4-spin cluster can assume five
states: -2, -1, 0, 1 and 2 (Eq.~\ref{totalspin}). The five-state
Kr\"onecker's delta reads
\begin{align}
  \nonumber \delta_{ss_\alpha} =1 - \frac{5}{4}\big(s^2 +
  s_\alpha^2\big) & + \frac{65}{72}ss_\alpha + \frac{1}{4}\big(s^4 +
  s_\alpha^4\big) - \frac{17}{72}\big(s^3s_\alpha +
  ss_\alpha^3\big)\\ & + \frac{707}{288}s^2 s_\alpha^2
  -\frac{155}{288}\big(s^4 s^2_\alpha + s^2 s_\alpha^4\big) +
  \frac{5}{72}s^3 s_\alpha^3 + \frac{35}{288}s^4
  s_\alpha^4\,. \label{delta5}
\end{align}
Introducing Eq. (\ref{delta5}) in Eq. (\ref{RS5}) and summing over $s$ results
\begin{align}
  \label{RS6} P(\mathbf{s}) & = \Big\langle \exp\beta\sum_\alpha
  \Big[- H_0(\mathbf{s}^\alpha{,}\{J_{ij}\}) + h_e s_\alpha\Big]
  \Big\rangle_{\{J_{ij}\}}\\ & \times \sum_k P_k \int \prod_{l=1}^k
  d\mathbf{h}_l\,W(\mathbf{h}_l) \Big\langle\exp\Big[ \beta
    \sum_l\boldsymbol{\phi}(\mathbf{h}_l{,}J_l{,}\{J_{ij}^l\})
    \cdot\sum_\alpha\mathbf{M}(s_\alpha)\Big]
  \Big\rangle_{\{J_l\},\{J_{ij}^l\}}\,, \nonumber
\end{align}
where $\boldsymbol{\phi}(\mathbf{h}_l{,}J_l{,}\{J_{ij}^l\})$ is a
vector with components
\begin{align}
  \phi_1(\mathbf{h}{,}J{,}\{J_{ij}\}) =
  \frac{1}{\beta}\Big[-\frac{1}{12}
    \ln\frac{\chi(2{,}\mathbf{h}{,}J{,}\{J_{ij}\})}
            {\chi(-2{,}\mathbf{h}{,}J{,}\{J_{ij}\})} + \frac{2}{3}
            \ln\frac{\chi(1{,}\mathbf{h}{,}J{,}\{J_{ij}\})}
                    {\chi(-1{,}\mathbf{h}{,}J{,}\{J_{ij}\})}\Big]\,,
\end{align}
\begin{align}
  \phi_2(\mathbf{h}{,}J{,}\{J_{ij}\}) & =
  \frac{1}{\beta}\Big[-\frac{1}{24}
    \ln\chi(2{,}\mathbf{h}{,}J{,}\{J_{ij}\})\chi(-2{,}\mathbf{h}{,}J{,}\{J_{ij}\})
    \\ & + \frac{2}{3} \ln\chi(1{,}\mathbf{h}{,}J{,}\{J_{ij}\})
    \chi(-1{,}\mathbf{h}{,}J{,}\{J_{ij}\}) -
    \frac{5}{4}\ln\chi(0{,}\mathbf{h}{,}J{,}\{J_{ij}\})\Big]\,,
  \nonumber
\end{align}
\begin{align}
  \phi_3(\mathbf{h}{,}J{,}\{J_{ij}\}) = \frac{1}{\beta}\Big[
    \frac{1}{12}
    \ln\frac{\chi(2{,}\mathbf{h}{,}J{,}\{J_{ij}\})}{\chi(-2{,}
      \mathbf{h}{,}J{,}\{J_{ij}\})} - \frac{1}{6}
    \ln\frac{\chi(1{,}\mathbf{h}{,}J{,}\{J_{ij}\})}{\chi(-1{,}
      \mathbf{h}{,}J{,}\{J_{ij}\})}\Big]
\end{align}
and
\begin{align}
  \phi_4(\mathbf{h}{,}J{,}\{J_{ij}\}) & =
  \frac{1}{\beta}\Big[\frac{1}{24}
    \ln\chi(2{,}\mathbf{h}{,}J{,}\{J_{ij}\})\chi(-2{,}
    \mathbf{h}{,}J{,}\{J_{ij}\}) \\ & - \frac{1}{6}
    \ln\chi(1{,}\mathbf{h}{,}J{,}\{J_{ij}\})\chi(-1{,}
    \mathbf{h}{,}J{,}\{J_{ij}\}) + \frac{1}{4}\ln\chi(0{,}
    \mathbf{h}{,}J{,}\{J_{ij}\})\Big]\,. \nonumber
\end{align}
Comparing Eq. (\ref{RS6}) with the RS {\sl Ansatz},
Eq. (\ref{RSansatz}), in the limit $n\rightarrow 0$ we have
\begin{align}
  \nonumber \int d\mathbf{h}\, W(\mathbf{h}) & \Big\langle\exp\beta\sum_\alpha
  \Big[-H_0(\mathbf{s}^\alpha{,}\{J_{ij}\}) + h_e s_\alpha +
    \mathbf{h} \cdot \mathbf{M}(s_\alpha)\Big] \Big\rangle_{\{J_{ij}\}} \\
  \label{RS7} & = \Big\langle  \exp\beta\sum_\alpha
  \Big[- H_0(\mathbf{s}^\alpha{,}\{J_{ij}\}) + h_e s_\alpha\Big]
  \Big\rangle_{\{J_{ij}\}}\\ \times \sum_kP_k \int & \prod_{l=1}^k
  d\mathbf{h}_l\,W(\mathbf{h}_l) \Big\langle \exp\Big[
    \beta\sum_l\boldsymbol{\phi}(\mathbf{h}_l{,}J_l{,}\{J_{ij}^l\})
    \cdot\sum_\alpha\mathbf{M}(s_\alpha)\Big]
  \Big\rangle_{\{J_l\},\{J_{ij}^l\}}\,.  \nonumber
\end{align}
Introducing a Dirac's $\delta$-function for each field component in the RHS results
\begin{align}
  \nonumber \int d\mathbf{h}\, & W(\mathbf{h})
  \Big\langle\exp\beta\sum_\alpha
  \Big[-H_0(\mathbf{s}^\alpha{,}\{J_{ij}\}) + h_e s_\alpha +
    \mathbf{h} \cdot \mathbf{M}(s_\alpha)\Big]
  \Big\rangle_{\{J_{ij}\}} \\ = \Big\langle &
  \exp\beta\sum_\alpha\Big[- H_0(\mathbf{s}^\alpha{,}\{J_{ij}\}) + h_e
    s_\alpha\Big] \Big\rangle_{\{J_{ij}\}} \nonumber \int
  d\mathbf{h}\, \sum_kP_k \int\prod_{l=1}^k
  d\mathbf{h}_l\,W(\mathbf{h}_l) \\ \times &
  \Big\langle\prod_{i=1}^4\delta\Big(h_i -
  \sum_l\phi_i(\mathbf{h}_l{,}J_l{,}\{J_{ij}^l\})\Big) \exp\Big[
    \beta\mathbf{h} \cdot\sum_\alpha \mathbf{M}(s_\alpha)\Big]
  \Big\rangle_{\{J_l\},\{J_{ij}^l\}}\,. \label{RS8}
\end{align}

Comparing both sides, we obtain the saddle-point equation for the RS
distribution of local fields, Eq. (\ref{RS9}).

\bibliographystyle{elsarticle-num} 
\bibliography{bibliografia}

\end{document}